\documentclass[10pt, reqno, oneside]{article}

\date{April 24, 2013}
\author{D{\'a}vid Papp \and Jan Unkelbach}
\title{Direct leaf trajectory optimization for volumetric modulated arc therapy planning with sliding window delivery}

\usepackage{amsmath}
\usepackage[margin=1.25in]{geometry}
\usepackage{graphicx}
\usepackage{booktabs}
\usepackage{multirow}
\usepackage{url}

\begin{document}

\maketitle


\abstract{We propose a novel optimization model for volumetric modulated arc therapy (VMAT) planning that directly optimizes deliverable leaf trajectories in the treatment plan optimization problem, and eliminates the need for a separate arc-sequencing step. In this model, a 360-degree arc is divided into a given number of arc segments in which the leaves move unidirectionally. This facilitates an algorithm that determines the optimal piecewise linear leaf trajectories for each arc segment, which are deliverable in a given treatment time. Multi-leaf collimator (MLC) constraints, including maximum leaf speed and interdigitation, are accounted for explicitly. The algorithm is customized to allow for VMAT delivery using constant gantry speed and dose rate, however, the algorithm generalizes to variable gantry speed if beneficial.
We demonstrate the method for three different tumor sites: a head-and-neck case, a prostate case, and a paraspinal case. For that purpose, we first obtain a reference plan for intensity modulated radiotherapy (IMRT) using fluence map optimization and 20 equally spaced beam directions. Subsequently, VMAT plans are optimized by dividing the 360-degree arc into 20 corresponding arc segments. Assuming typical machine parameters (a dose rate of 600 MU/min, and a maximum leaf speed of 3 cm/sec), it is demonstrated that the quality of the optimized VMAT plans approaches the quality of the IMRT benchmark plan for delivery times between 3 and 4 minutes.}

\setlength{\parskip}{0.5em plus 0.2em minus 0.2em}

\section{Introduction}\label{sec:introduction}
Radiotherapy delivery using volumetric modulated arc therapy (VMAT) \cite{yu2011intensity} has gained popularity since it bears the potential to improve treatment plan quality while shortening treatment time. The potential for improved plan quality originates from the fact that all beam directions in a $360^\circ$ arc are utilized, whereas traditional IMRT treatments typically use 5-9 discrete beam angles. The potential to shorten treatment times stems from the feature that the treatment beam is on continuously while the gantry and the MLC leaves are in motion.

Over the past years, treatment plan optimization methods for IMRT planning have evolved to a high level of sophistication. One major advantage of IMRT planning is that the fluence map optimization (FMO) problem can be formulated as a convex optimization problem, which can be solved to near optimality using established algorithms. In addition, direct aperture optimization (DAO) methods \cite{shepard02,romeijn05} have been developed for step \& shoot delivery, which can closely approximate the FMO solution with a small number of apertures \cite{carlsson2008combining,cassioli13}.

The goal of VMAT planning is to find the optimal trajectories of the leaves of a multi-leaf collimator, which yield the best plan quality for a given treatment time. In contrast to the FMO problem, the general VMAT planning problem cannot be formulated as a continuous convex optimization problem.  Therefore, treatment plan optimization for VMAT still faces substantial challenges.

One approach to VMAT planning is a two-step approach in analogy with the traditional IMRT planning. In the first step, an FMO problem is solved for a relatively large number of equispaced beam angles. In the second step, an arc-sequencing method is used to convert the FMO solution into deliverable arcs \cite{craft09,wang2008arc}. VMAT planning approaches in this category differ in the way arc-sequencing is performed. In the algorithm proposed by Craft, \cite{craft12}, FMO is performed for 180 equispaced beam directions. Sequencing is performed by iteratively merging similar fluence maps of neighboring beam angles. The merged fluence maps are finally delivered over the corresponding arc segment using a sliding window technique. The work by Wang \textit{et al.} \cite{wang2008arc,wang2011CCPP} suggests an arc sequencing approach based on shortest path algorithms. Their algorithm starts with a coarser FMO solution (20 equispaced beams) which are delivered over a corresponding arc segment. Leaf trajectories are obtained by assuming unidirectional motion of a leaf within each segment. The algorithm of Cameron \cite{Cameron2005sweeping} attempts to optimize the leaf trajectories using simulated annealing, starting from back-and-forth sweeping initial trajectories consisting of 3--7 segments.

Alternative approaches to VMAT planning can be considered as extensions of direct aperture optimization methods and differ in the approach to DAO that is used. In the algorithm presented by Bzdusek\footnote{which is the basis for the commercial implementation SmartArc in the Pinnacle planning system} \cite{bzdusek09}, a local gradient based method to leaf position refinement is used, starting from initial apertures derived from a course FMO solution. Ulrich \cite{ulrich2007development} suggests a leaf position optimization method based on tabu search. In the approach of Otto\footnote{which is the basis for the commercial implementation RapidArc in the Eclipse planning system} \cite{otto08}, stochastic search methods are used to tweak leaf positions. The approach by Peng \cite{peng12} extends the column generation approach to DAO to the VMAT planning problem. In the column generation approach, apertures are added to a treatment plan one after another until one aperture per beam angle is generated. New apertures are selected by solving a so-called pricing problem, in which the aperture that promises the largest improvement to the plan quality is identified.


In this paper, we propose a new algorithm to VMAT planning, which directly optimizes the leaf trajectories of a multi-leaf collimator. Key advantages of the proposed algorithm are:
\begin{itemize}
\item[$\bullet$] The method generates a deliverable VMAT plan in a single step without the need for an arc-sequencing step. The total delivery time is specified by the treatment planner and is met exactly by the algorithm, and so are the machine constraints.
\item[$\bullet$] When allowing for enough delivery time, the plan quality converges to the ideal FMO solution. For the cases presented in this paper, convergence was observed after 3-4 minutes.
\item[$\bullet$] The algorithm facilitates VMAT delivery with constant gantry speed and dose rate, which may be preferable from a delivery perspective. However, it generalizes to variable gantry speed if beneficial for a patient geometry. 
\end{itemize}

The remainder of this paper is organized as follows: In section \ref{sec:solution} we derive the VMAT planning algorithm. In section \ref{sec:results} we present results for a head \& neck case, a paraspinal case, and a prostate case. In section \ref{sec:discussion} we discuss the results.

%

\section{VMAT planning approach}\label{sec:solution}

In this section, we first outline the key ideas behind our solution approach (Section \ref{sec:solution-intro}), and then formulate our basic treatment planning optimization model for direct leaf trajectory optimization with sliding window delivery (Section \ref{sec:model}). This model is represented as a convex optimization problem that can be solved using readily available numerical optimization software. In section \ref{sec:dose} we present a method to improve dose calculation accuracy over the basic model of section \ref{sec:model}.  Finally, \mbox{Section \ref{sec:extensions}} is concerned with extensions that can be added to the model without increasing the complexity of the treatment planning procedure, but are not considered in the previous sections, in order to keep the presentation simple.

\subsection{Intuition behind the approach}\label{sec:solution-intro}

Our approach is inspired by the following observations: First, it is mostly agreed upon that IMRT plans using a large enough number of equispaced beam angles (in this paper we use 20) can very closely approximate the ideal dose distribution obtained from a 180 beam IMRT plan \cite{bortfeld2010number}. Second, every fluence map of an IMRT plan can be delivered accurately using a sliding window technique, where the leaves move unidirectionally across the field. This motivates the following approach to VMAT planning: A 360-degree arc is divided into a fixed number of arc segments (in this paper we use 20). In each arc segment, the leaves move unidirectionally, thus delivering an effective fluence map over each arc segment. Allowing for enough delivery time, an arbitrary effective fluence map can be generated, thus allowing for a close approximation of a 20 beam IMRT plan. The approach has a number of key advantages:
\begin{itemize}
\item[$\bullet$] The restriction imposed on the leaf trajectories (i.e. unidirectional within each arc segment) allows us to directly optimize the leaf trajectories along with finding the optimal dose distribution. The leaf trajectory optimization problem can be formulated as a continuous optimization problem, which is convex under certain assumptions. Thus, near optimal solutions can be found reliably using convex optimization methods.
\item[$\bullet$] Instead of pursuing a two-step approach in which fluence maps are obtained first, which are subsequently converted into a leaf trajectory, we directly optimize piecewise linear leaf trajectories. This allows us to take MLC constraints and delivery time explicitly into account during plan optimization. In addition, a separate arc sequencing step, which may compromise treatment quality, is avoided.
\item[$\bullet$] Given sufficient delivery time, the optimal VMAT plan can perfectly match the ideal IMRT plan. On the other hand, we can still gain delivery time over a DMLC delivery of the ideal IMRT plan, as it is expected from a VMAT plan.
\end{itemize}

\subsection{The direct leaf trajectory optimization model}\label{sec:model}

The direct leaf trajectory optimization model is an extension of the standard fluence map optimization (FMO) model,
\begin{equation}\tag{FMO}\label{eq:FMO}
\begin{aligned}
\text{minimize}\quad   & f(d)\\
\text{subject to}\quad & d_i = \sum_{k=1}^K\sum_{n=1}^N\sum_{j=1}^J D^{k}_{nij} x^{k}_{nj} & \forall\,i\\
                       & x^{k}_{nj} \geq 0 & \forall\,k,n,j
\end{aligned}
\end{equation}
where we optimize a convex objective function $f$ with respect to the variables $x$ and $d$. Here, $d_i$ is the dose absorbed by voxel $i$.  The superscript $k$ is the index of the beam angle; the subscripts $n$ and $j$ are the row and column indices in the bixel map, i.e. $n$ is the index of the MLC leaf pair, and $j$ is the index of the bixel in leaf motion direction. Thus, $x^k_{nj}$ is the fluence corresponding to bixel $j$ of the MLC leaf pair $n$ in beam $k$. The dose deposition coefficients $D^{k}_{nij}$ denote the dose contributions of bixel $(k,n,j)$ to voxel $i$ for unit fluence.

Rather than optimizing the fluence $x$, and then determining a leaf trajectory that delivers it approximately within a given time, we include in the model the leaf trajectories of an optimal sliding window delivery, which is determined concurrently with the optimal fluence. To that end, we divide the 360-degree arc into $K$ arc segments in which the leaves move unidirectionally. From now on, we thus associate the index $k$ with an arc segment, rather than a discrete beam angle. We assume for now that the dose-influence matrix is constant over the arc segment, that is, $D^k_{nij}$ depends on $k$, but does not change with the gantry angle within segment $k$. In each segment $k$ an effective fluence map is delivered, whose fluences will also be denoted by $x^k_{nj}$.

\begin{figure}
\centering
\includegraphics[width=250pt]{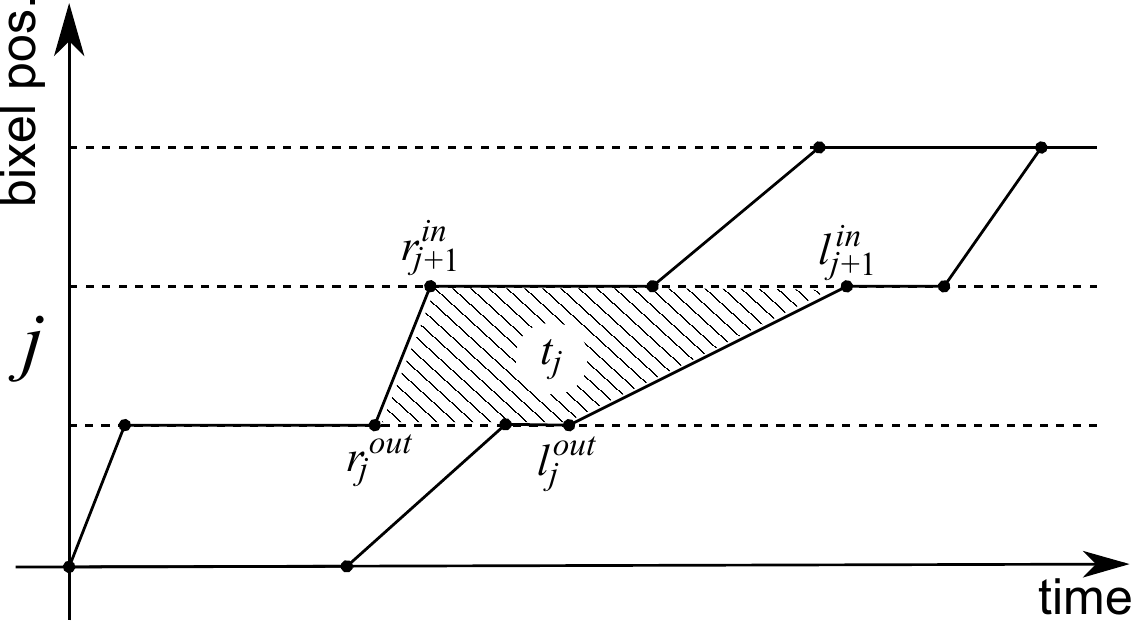}
\caption{Leaf trajectories and ``effective beam-on time'' for a single leaf pair $n$ in a single arc segment $k$. The effective beam-on time is the shaded area between the leaf trajectories and the bixel boundaries. The arc segment index $k$ and leaf pair index $n$ are dropped from the leaf position variables for readability.}\label{fig:eff-beam-on}
\end{figure}

Consider a leaf pair $n$ as it moves across the field in arc segment $k$ as the gantry rotates at a constant speed. The constant gantry speed defines a linear one-to-one mapping between time and gantry angles, and the leaf positions can be expressed as functions of time or gantry angle. Let $r_{knj}^{in}$ and $r_{knj}^{out}$ denote the time the leading leaf of leaf pair $n$ reaches and leaves bixel $j$ as it sweeps across the field, i.e. at time $r_{knj}^{in}$ it reaches the boundary between bixels $j-1$ and $j$, and at time $r_{knj}^{out}$ it leaves the boundary between bixels $j-1$ and $j$. Analogously, let $\ell_{knj}^{in}$ and $\ell_{knj}^{out}$ denote the same for the trailing leaf. This notation is illustrated in Fig.~\ref{fig:eff-beam-on}: It can be seen that the \emph{effective beam-on time} for bixel $j$ is given by 
\[ t^k_{nj} = \frac{1}{2}\left[ (\ell_{knj}^{out}-r_{knj}^{out}) + (\ell_{kn(j+1)}^{in}-r_{kn(j+1)}^{in})\right], \]
which corresponds to the area between the trajectories of the leading and trailing leaves. Assuming a fixed dose rate $\delta$ throughout the delivery and a constant dose-influence matrix $D^k$ for each arc segment $k=1,\dots,K$ leads to the following formulation:

\begin{equation}\label{eq:VMAT-1}
\begin{aligned}
\text{minimize}\quad   & f(d)\\
\text{subject to}\quad & d_i = \sum_{k=1}^K\sum_{n=1}^N\sum_{j=1}^J D^{k}_{nij} x^{k}_{nj} \qquad \forall\,i\\
                       & x^k_{nj} = \delta\cdot t^{k}_{nj} \qquad \forall\,k,n,j \\
                       & t^k_{nj} = (\ell_{knj}^{out}-r_{knj}^{out} + \ell_{kn(j+1)}^{in}-r_{kn(j+1)}^{in})/2 \qquad \forall\,k,n,j\\
                       & (\ell,r) \text{ encode an feasible leaf trajectory}.
\end{aligned}
\end{equation}


To complete our model, we need to specify the ``feasible leaf trajectories'' of \eqref{eq:VMAT-1} in terms of the leaf arrival and departure times. We need to ensure that the order of the leading and trailing leaves are preserved, the leaves move in the right direction, and that the leaves do not travel faster than the collimator's maximum leaf speed. Moreover, the total delivery time must be below the maximum allotted time. The precise algebraic form of some of the constraints depends on whether the leaves move from left to right or from right to left; below we give the complete formulation for the left-to-right direction, and indicate the necessary changes for the opposite direction.

Let us denote the maximum total delivery time by $t_{\max}$. The maximum leaf speed and the bixel size determine the minimum time $\Delta t$ a leaf needs to traverse a bixel. With these notation, feasible leaf trajectories are those in which every leaf pair satisfies the constraints
\begin{subequations}\label{eq:VMAT-2}
\begin{alignat}{2}
&0 \leq r_{knj}^{in} \leq r_{knj}^{out} \leq t_{\max}/K,\quad && 0 \leq \ell_{knj}^{in} \leq \ell_{knj}^{out} \leq t_{\max}/K, \label{eq:VMAT-2a}\\
&r_{knj}^{out}+\Delta t \leq r_{kn(j+1)}^{in},           && \ell_{knj}^{out}+\Delta t \leq \ell_{kn(j+1)}^{in},\label{eq:VMAT-2b}\\
&r_{knj}^{in} \leq \ell_{knj}^{in},                    && r_{knj}^{out} \leq \ell_{knj}^{out}\label{eq:VMAT-2c}
\end{alignat}
\end{subequations}
for every $j$ and $n$ in those arc segments $k$ where leaves move from left to right. The inequalities \eqref{eq:VMAT-2a} ensure that the breakpoints in the piecewise linear leaf trajectories are properly ordered, and that the plan is delivered in the allotted time; \eqref{eq:VMAT-2b} implements the maximum leaf speed constraint imposed by the MLC; and \eqref{eq:VMAT-2c} keeps the trailing leaf behind the leading one. In the arc segments where the leaves move in the right-to-left direction, the constraints are identical to the ones above except that the subscripts $j+1$ in \eqref{eq:VMAT-2b} need to be replaced with $j-1$.

Note that all the constraints in \eqref{eq:VMAT-1} and \eqref{eq:VMAT-2} are linear in the decision variables. Therefore, if we assume that the dose-influence matrix is constant within each arc segment, then the direct leaf trajectory optimization problem is a convex optimization problem with linear constraints. As such, it is solvable to global optimality with readily available convex optimization methods.

\subsection{Improved dose calculation with variable dose-influence matrices}\label{sec:improved-model}
\label{sec:dose}
The above model is only approximate, because we assumed the dose-influence matrices $D$ to be constant over each arc segment. This will not provide a sufficiently accurate dose calculation for the VMAT plan. To increase the dose calculation accuracy, we calculate dose influence matrices $D$ at a finer angular resolution. For the results presented in this paper, we calculate 180 dose influence matrices at a $2$ degree resolution. In contrast, the 360 degree arc is divided into 20 segments; thus every arc segment is subdivided into 9 sectors with distinct dose influence matrices. To account for this in the optimization model, we need to replace in \eqref{eq:VMAT-1} the arc segment index $k$ in the superscript of $D$ with an angle index $a(k,n,j)$ at which bixel $j$ of the $n$th leaf pair is open in the $k$th arc segment. 


From a computational perspective the more accurate model is more difficult, because the dose deposition coefficients' dependence on the leaf trajectories renders the optimization model non-convex. We propose the following simple iterative method to mitigate this difficulty:
\begin{enumerate}
	\item Start with some initial values for the values of each $a(k,n,j)$. In our implementation the initial angle $a(k,n,j)$ was chosen to correspond to the middle of the arc segment $k$ for every $n$ and $j$.
	\item Solve the (convex) model \eqref{eq:VMAT-1}-\eqref{eq:VMAT-2} with the current values of $a(k,n,j)$.
	\item Using the obtained leaf trajectories, determine the beam angles at the times the leaves traverse each bixel $j$, and update the values $a(k,n,j)$ accordingly; then return to step 2.
\end{enumerate}

Ideally, we iterate in the above procedure until no $a(k,n,j)$ is updated; in practice, we can stop when the updates are small enough. In our implementation, we stopped whenever no more than $5\%$ of the $a(k,n,j)$ changed, and none of them changed by more than $2$ degrees. We found that in most cases only a very few (2 or 3) update steps were needed before we could stop.

\begin{figure}
\centering
\includegraphics[width=300pt]{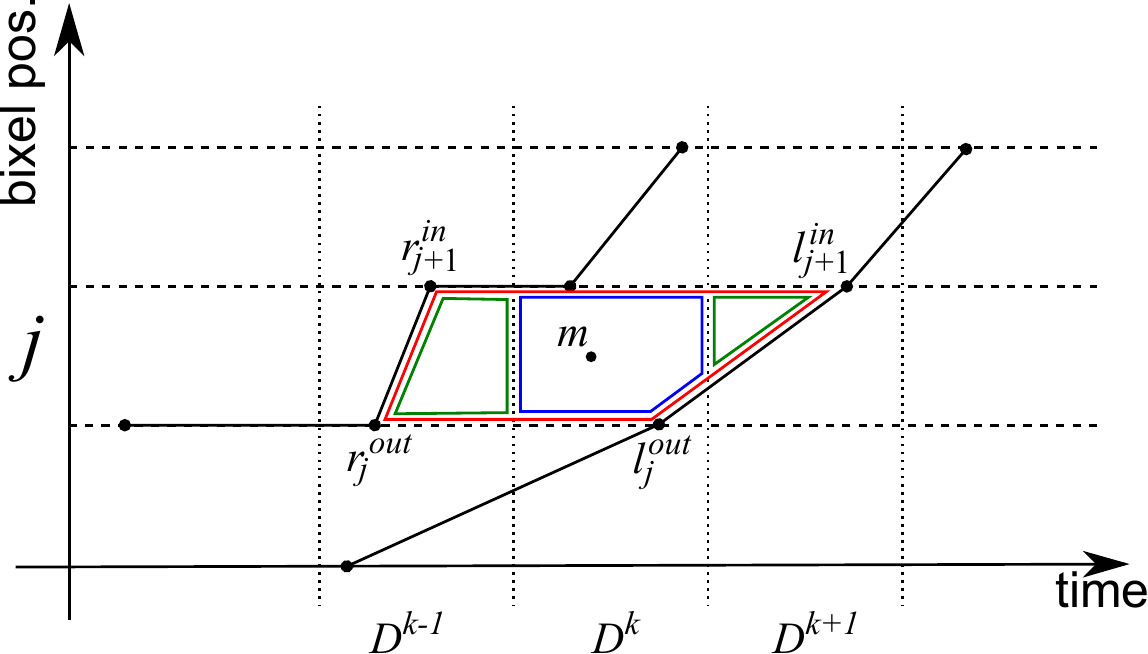}
\caption{Determining the ``effective angle'' at which a bixel is open, and the dose distribution. The corners of the effective beam-on interval (red quadrangle) determine the time point $m = (r_j^{out}+\ell_j^{out}+r_{j+1}^{in}+\ell_{j+1}^{in})/4$; the corresponding angle is the effective angle. The area of this quadrangle can be used as an approximate beam-on time of bixel $j$, in connection with the dose-influence matrix $D^k$ for dose calculation. A more precise dose calculation is possible by partitioning the effective beam-on interval into subintervals corresponding to the sectors of the dose-influence matrices (green and blue areas). To simplify the formulation, the approximate formula was used throughout our VMAT optimization algorithm, but the final dose calculation was done using the exact method.}\label{fig:midpoint}
\end{figure}

There are also multiple ways to implement the update step, owing to the fact that the gantry moves slightly as the leaves traverse bixel $j$, and a dose-influence matrix corresponding to any of the corresponding angles might be chosen. This situation is schematically illustrated in Fig.~\ref{fig:midpoint}. During optimization, we assign each bixel $(k,n,j)$ to a single dose-influence matrix. More specifically, we choose the angle index $a(k,n,j)$ such that it corresponds to the midpoint of bixel $j$'s effective beam-on time interval, which is given by $(r_j^{out}+\ell_j^{out}+r_{j+1}^{in}+\ell_{j+1}^{in})/4$. (See Fig.~\ref{fig:midpoint}.) In most cases this is sufficient, since (assuming a $2$ degree resolution of the dose-influence matrices) there are more bixels to traverse in an arc segment than there are dose-influence matrices, and the leaves do not get enough time to stay within the same bixel for longer than the gantry rotates a few ($2$--$4$) degrees. Therefore, it is expected that (especially for shorter treatment times) the dose-influence matrix is approximately constant during the short time interval that a leaf pair exposes a particular bixel. 

\paragraph{Final dose calculation step:} Even though a bixel can typically be assigned to a single dose-influence matrix, we perform a more accurate final dose calculation step after the optimization. To that end, we proportionally assign the effective beam-on-time for each bixel to multiple adjacent dose-influence matrices, which is also illustrated in Fig.~\ref{fig:midpoint}. However, in the numerical experiments, only minor differences were observed between the optimized and final dose distribution.

\subsection{Model extensions}\label{sec:extensions}

The optimization model presented in the previous section can be extended in a straightforward fashion to take into account additional machine constraints, and to be applicable in a more general setting. We outline a few of these possibilities in this section.

\subsubsection{Interdigitation}

Interdigitation constraints can be added to the model as additional linear constraints similar to those in Eq.~\eqref{eq:VMAT-2c}. To ensure that the trailing leaf of a leaf pair does not overlap with the leading leaf of an adjacent pair, it is sufficient to add the constraints
\[ r_{knj}^{in} \leq l_{k(n+1)j}^{in},\quad r_{knj}^{out} \leq l_{k(n+1)j}^{out}, \quad r_{knj}^{in} \leq l_{k(n-1)j}^{in},\quad r_{knj}^{out} \leq l_{k(n-1)j}^{out} \qquad \forall\,k,n,j.\]
Since only a small number of linear constraints are added, the additional effort needed to solve this extended model is not substantial.

\subsubsection{Varying gantry speed}
\label{sec:gantry}
The model \eqref{eq:VMAT-1}-\eqref{eq:VMAT-2} optimizes delivery for a given constant gantry speed. Although this may be desirable from a delivery perspective\footnote{It is expected to result in a more reliable translation from the treatment plan to actual machine parameters and reduced machine wear.}, it might be suboptimal if the allotted treatment time is better spent in some arc segments than in others. Our model can be extended to the case when the gantry rotates at a constant speed within each arc segment, but this speed is allowed to be different for each arc segment. Our model can optimize for this piecewise constant gantry speed in the following way:

Let us introduce a new decision variable $t_{\max}^k$ for each arc segment $k$ ($k=1,\dots,K$), and denote the maximum total delivery time by $t_{\max}$ as before. Then drop from the constraints \eqref{eq:VMAT-2a} the inequalities involving $t_{\max}$, and replace them with
\[ r_{knj}^{out} \leq t^k_{\max}, \quad \ell_{knj}^{out} \leq t^k_{\max} \]
to limit the leaf travel times within each arc segment; finally, add one last constraint to bound the total treatment time:
\[ \sum_{k=1}^K t^k_{\max} \leq t_{\max}. \]
This extension does not change the complexity of the optimization model, as it only requires the addition of a small number of variables and linear constraints.

\subsubsection{Multiple arcs and partial arcs}

So far we have assumed that the treatment is delivered in a single 360-degree arc. Naturally, nothing in the model relies on the 360-degree arc assumption, the same approach can be used verbatim to optimize the leaf trajectories within a partial arc with given initial and terminal angle. Note, however, that the leaves need enough time within each arc segment for both modulation and crossing the entire field. Hence, in order to realize a delivery time reduction, partial arcs should be divided into proportionately fewer arc segments than complete arcs\footnote{As opposed to reducing the length of the arc segments while keeping the number of arc segments constant.}. 

Multiple-arc treatments can be optimized with our approach essentially verbatim. Leaf trajectories can be encoded separately for each arc, and the dose calculation constraint in \eqref{eq:VMAT-1} can be changed to a quadruple sum that computes the total dose in all the arcs. This might be beneficial even for sliding window delivery, for instance, in the case of large treatment sites, and plans requiring multiple arcs with difference isocenters.

\section{Computational results}\label{sec:results}



In this section we demonstrate the performance of our method for three different treatment sites: 1) a head \& neck case for which the lymph nodes are part of the CTV; 2) a prostate case for which the seminal vesicles are treated; and 3) a paraspinal case for which the tumor entirely surrounds the spinal cord and is located in proximity to the kidneys. We provide a detailed description of our experimental results for the head-and-neck case, and a summary of the prostate and paraspinal cases.

For each case, an IMRT plan was optimized using $20$ equispaced beam directions. The 20-beam IMRT plan, obtained through fluence map optimization (FMO) without applying any sequencing, was used as the reference plan. This reference plan was compared to single-arc VMAT plans, which were optimized for different treatment times ranging from 2 minutes to 5 minutes, using the same objective function. All VMAT plans were optimized for $20$ arc segments, constant gantry speed, and constant dose rate. The machine parameters in all the cases were given by a dose rate of $\delta = 600 \text{ MU/min}$ and a maximum collimator leaf speed of $3 \text{ cm/s}$, yielding $\Delta t = 1/3\text{ s}$ in the leaf trajectory constraints \eqref{eq:VMAT-2}.

Dose deposition coefficients for the VMAT planning were computed for $180$ equispaced \mbox{$6$ $\text{MV}$} beams with \mbox{$1 \times 1$ $\text{cm}^2$} beamlet grid resolution, using the quadratic infinite beam (QIB) dose-calculation algorithm implemented in \mbox{CERR 4.0b4}; see \cite{deasy2003cerr}. The dose deposition coefficients obtained from CERR are converted into the natural units Gy/MU, such that the effective bixel intensities are given in MU and the dose is obtained in Gy. The maximum field size was adapted to each case. We determined the smallest rectangular field that covers all target voxels from all of the 180 beam angles. The voxel resolution also varied from case to case, and was given by \mbox{$2.3 \times 2.3 \times 1.25$ $\text{mm}^3$} for the paraspinal case, \mbox{$3.9 \times 3.9 \times 2.5$ $\text{mm}^3$} for the prostate and head-and-neck cases. The collimator orientation was chosen such that the leaves move within the plane of the gantry rotation.


The objective function in our optimization model is a standard voxel-by-voxel piecewise quadratic penalty function that attempts to limit the over- and underdosing of various organs:
\begin{equation*}
\sum_i \left(w_i^+ \max(0,d_i - d_i^\text{hi})^2 + w_i^- \max(0,d_i^\text{lo} - d_i)^2\right),
\end{equation*}
where $d_i$ is the dose absorbed by voxel $i$, $d_i^\text{lo}$ and $d_i^\text{hi}$ are prescribed lower and upper bounds on this dose, and the coefficients $w_i^+$ and $w_i^-$ are treatment planning parameters. These coefficients were chosen case-by-case to obtain high quality IMRT plans. For each patient, the objective function parameters were kept the same for the IMRT plan and all VMAT plans. All treatment plans were optimized for a standard fractionated treatment, i.e. a prescribed dose of approximately 2 Gy per fraction.


The different plans are compared using the dose-volume histograms, colorwash dose distributions, and by computing their RTOG conformity indices and their conformity numbers proposed by van't Riet \textit{et al.} \cite{feuvret2006conformity}. These quantities are defined as
\[ \text{CI}_{RI} = \frac{V_{RI}}{TV} \quad \text{ and } \quad \text{CN}_{RI} = \frac{TV_{RI}}{TV}\cdot\frac{TV_{RI}}{V_{RI}}, \]
where 
$RI$ stands for reference isodose, $TV_{RI}$ = target volume covered by the reference isodose, $TV = $ total target volume, $V_{RI} = $ total volume covered by the reference isodose. The second quantity measures simulteneously the level of target coverage and the sparing of healthy tissues. The reference isodose is chosen as 95\% of the target prescription dose.

\subsection{Head-and-neck}

\begin{figure}
\centering
\includegraphics[height=120pt]{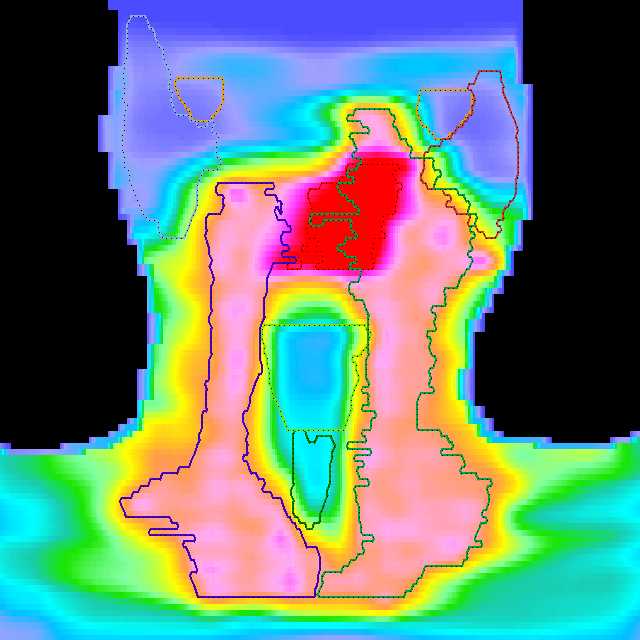}\quad
\includegraphics[height=120pt]{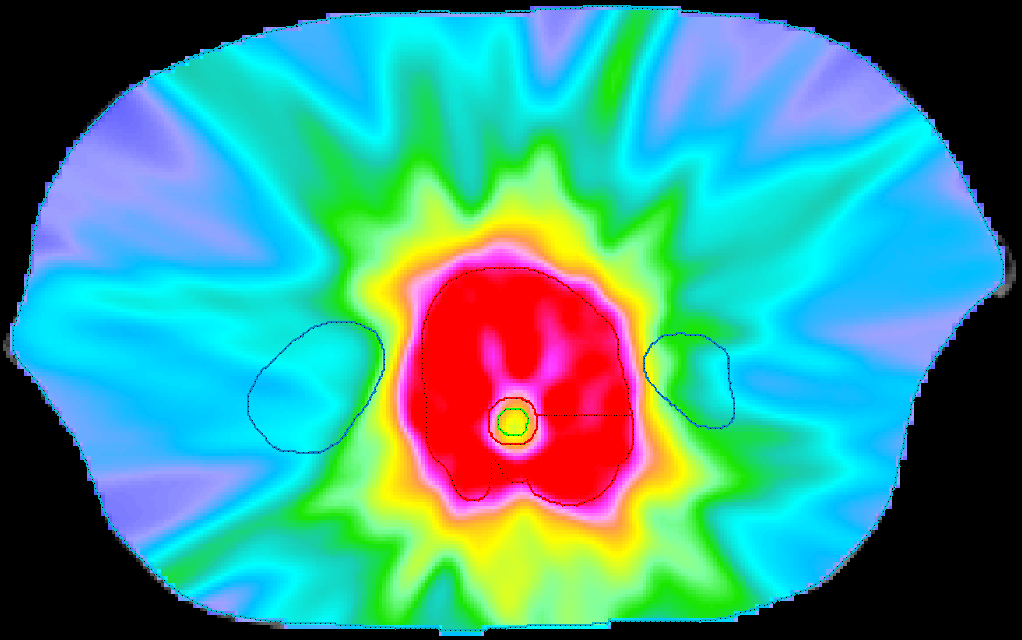}
\caption{Representative CT slices exhibiting the target geometry for the head-and-neck and the paraspinal case. In the head-and-neck case the secondary targets are the lymph nodes, their main dose-limiting organ is the trachea. In the paraspinal case the tumor wraps around the lower section of the spinal cord; the secondary dose limiting organs are the kidneys.}
\label{fig:hn-plus-paraspinal-colorwashes}
\end{figure}

The head-and-neck case includes three separate target volumes (PTVs): the primary tumor, and the right and left lymph nodes. The prescription was to deliver 66 Gy to the primary target and 54 Gy to the lymph nodes in 33 fractions. Eight organs at risk (OARs) were considered in treatment planning, with dose limits established in the AAPM planning guidelines for head-and-neck IMRT patients \cite{ezzell2003IMRTguidelines}. The main dose-limiting OARs were the parotid glands, the trachea, and the mandible. A representative axial CT slice through the superior part of the target is shown in Fig.~\ref{fig:head-and-neck-colorwash}; Fig.~\ref{fig:hn-plus-paraspinal-colorwashes} (left) shows a coronal view to illustrate the target geometry in three dimensions.

\begin{figure}
\centering
\includegraphics[width=400pt]{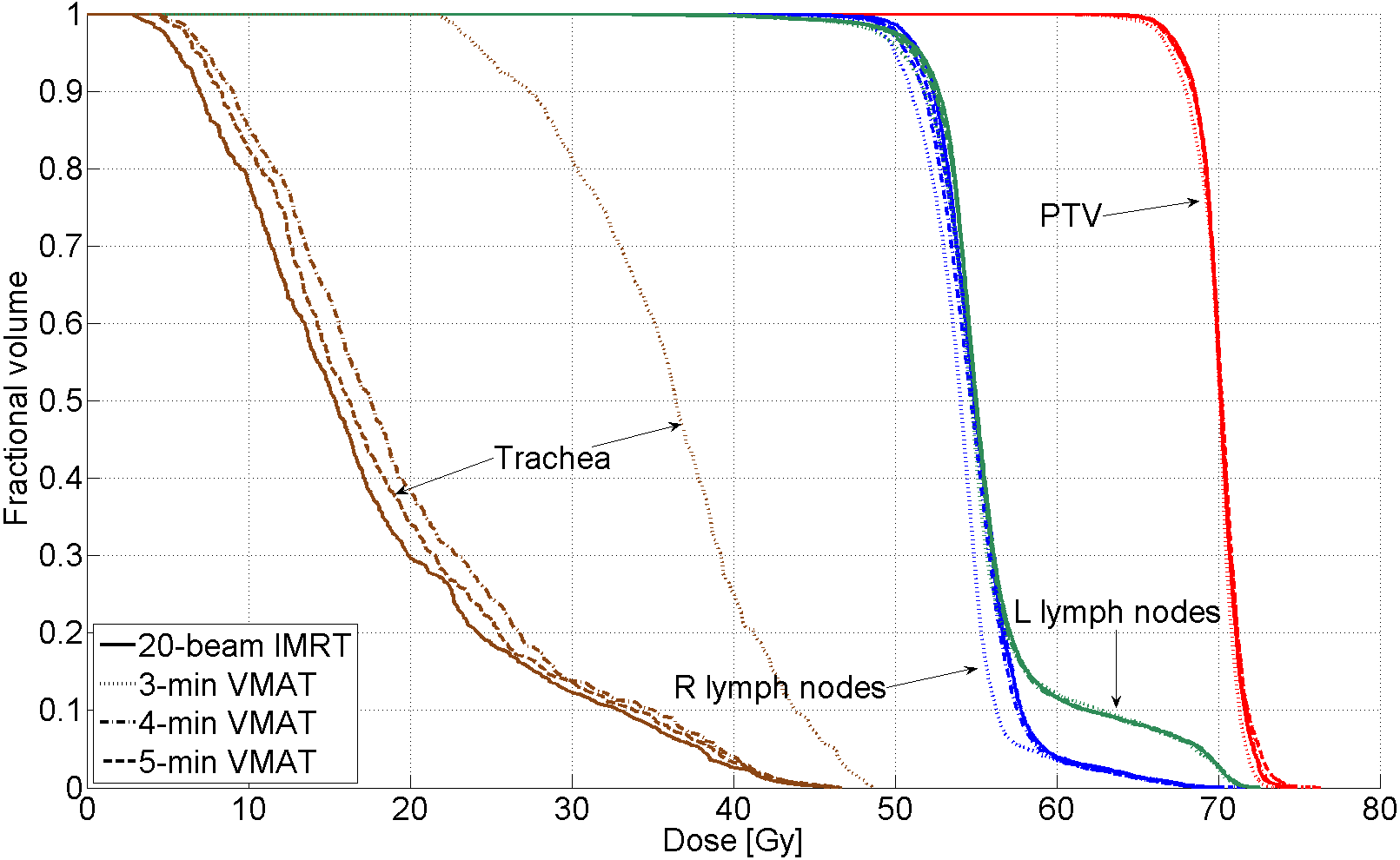}\\
\includegraphics[width=400pt]{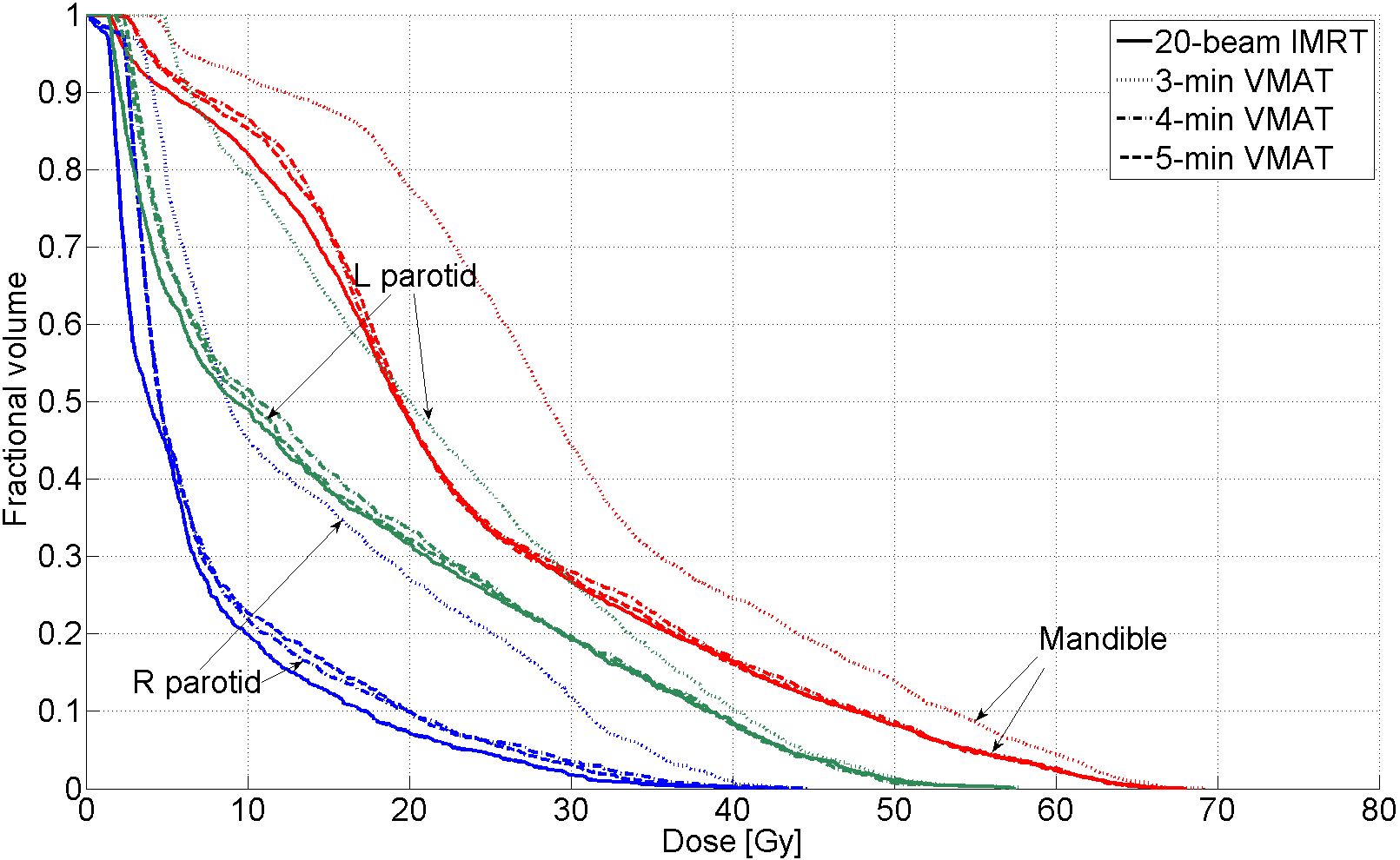}
\caption{Dose-volume histograms for the head-and-neck case comparing an optimal 20-beam IMRT plan and the VMAT plans with treatment time limit $t_{\max}=3$, $4$, and $5$ minutes. The diagram is split into two for better legibility. Only the target volumes (first panel) and the dose-limiting organs (trachea on first panel; parotid glands, mandible on second panel) are shown. The 4-minute and 5-minute plans are comparable to the optimal 20-beam IMRT plan both in their target coverage and organ sparing.}\label{fig:head-and-neck-DVH}
\end{figure}

We determined the optimal VMAT plans with our approach for the time limits $t_{\max} = 3, 4, 5$, and $20$ minutes, and compared their treatment quality to that of the optimal 20-beam FMO solution. The dose volume histograms (DVH) of the PTVs and the dose-limiting OARs is shown in Fig.~\ref{fig:head-and-neck-DVH}; the conformity indices and the conformation numbers are shown on Tbl.~\ref{tbl:head-and-neck-conformity}. It is apparent that all treatment plans yield adequate target coverage and differ primarily in OAR sparing\footnote{Note that the left lymph-node PTV has significant overlap with the primary PTV, which is the reason for the high dose in part of the lymph node.}. The VMAT plan optimized for 20 minutes delivery time (not shown) is indistinguishable from the IMRT plan. It is further observed that the 4-minute and 5-minute plans are nearly identical to the FMO solution in terms of target coverage, and they also result in similar OAR sparing. This finding is confirmed in Fig.~\ref{fig:head-and-neck-colorwash}, which shows the colorwash dose distributions for a representative slice through the three targets. The 4-minute and 5-minute VMAT plans are not only similar in terms of the DVHs for targets and OARs, but are also similar regarding the conformity of the dose distribution.


Fig.~\ref{fig:head-and-neck-DVH} further reveals that the 3-minute VMAT plan is substantially worse than the 4-minute VMAT plan. In addition, Fig.~\ref{fig:head-and-neck-colorwash} shows that the dose distribution is less conformal and more washed out. Hence, increasing the treatment time from 3 minutes to 4 minutes yields substantial improvements in plan quality, whereas further delivery time increases only yield marginal improvements.

It is possible to explain the large difference between the treatment qualities of the 3- and 4-minute plans, and the considerably smaller gain from any additional treatment time. The smallest rectangular field that covers all target volumes for this case is $23 \text{ cm}$ wide. Therefore, it takes almost $8$ seconds for the leaves to traverse the field once. With 3 minutes delivery time and $20$ arc segments, only 9 seconds are spent in each arc segment. Since the leaves need almost 8 seconds for travel, only little more than 1 second is left to modulate the radiation field. Increasing the treatment time from 3 to 4 minutes means that 12 seconds are spent in each arc segment. This more than \emph{triples} the excess time that can be used for modulation. Beyond that, the gain from each additional minute is progressively smaller.




\begin{figure}
\centering
\includegraphics[width=120pt]{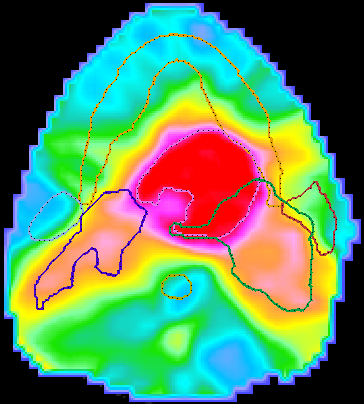}
\includegraphics[width=120pt]{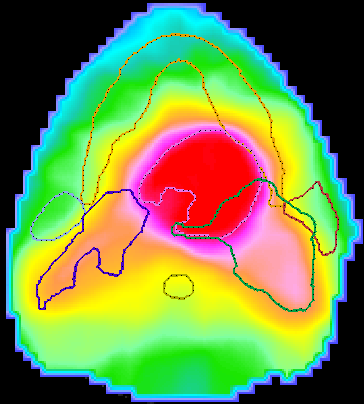}\\[0.3em]
\includegraphics[width=120pt]{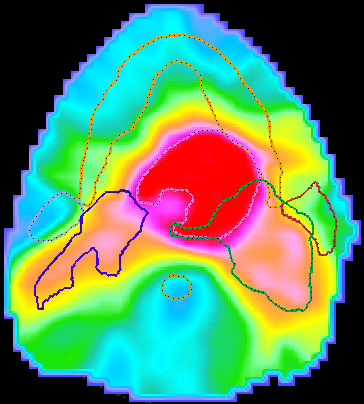}
\includegraphics[width=120pt]{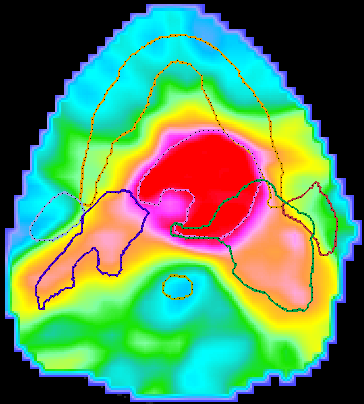}
\caption{Colorwash dose distributions on a representative CT slice for the 20-beam optimal IMRT plan (upper left), and the VMAT plans with treatment time limit $t_{\max}=3$ (upper right), $4$ (lower left), and $5$ (lower right) minutes. The contours of the spinal cord, mandible, parotid glands, lymph nodes, and the primary tumor are shown. While the 3-minute plan lacks both the conformity and the OAR sparing of the other plans, the 4- and 5-minute VMAT plans are nearly identical to the 20-beam IMRT plan.}
\label{fig:head-and-neck-colorwash}\label{fig:head-and-neck-colorwash}
\end{figure}

\begin{table}
\centering
\begin{tabular}{lcccc}
\toprule
      & 20-beam IMRT & 3 min VMAT & 4 min VMAT & 5 min VMAT \\
\midrule
$\text{CI}_{51.3}$ & 1.189 & 1.189 & 1.171 & 1.151 \\
$\text{CN}_{51.3}$ & 0.755 & 0.705 & 0.749 & 0.770 \\
\midrule
$\text{CI}_{62.7}$ & 1.420 & 1.502 & 1.420 & 1.420 \\
$\text{CN}_{62.7}$ & 0.702 & 0.664 & 0.701 & 0.701 \\
\bottomrule
\end{tabular}
\caption{The RTOG conformity indices and the van't Riet \textit{et al.} conformity numbers of the 20-beam IMRT plan and the optimal VMAT plans with different treatment time limits, corresponding to the 95\% of the 54 Gy and the 66 Gy target isodoses. The CI is better if it is closer to $1.0$, whereas the CN is better if it is higher. These indices show the essentially identical conformity of the 20-beam IMRT and the 4- and 5-minute plans. The 3-minute plan has noticeably lower conformity.}\label{tbl:head-and-neck-conformity}
\end{table}

\subsection{Prostate}

\begin{figure}
\centering
\includegraphics[width=400pt]{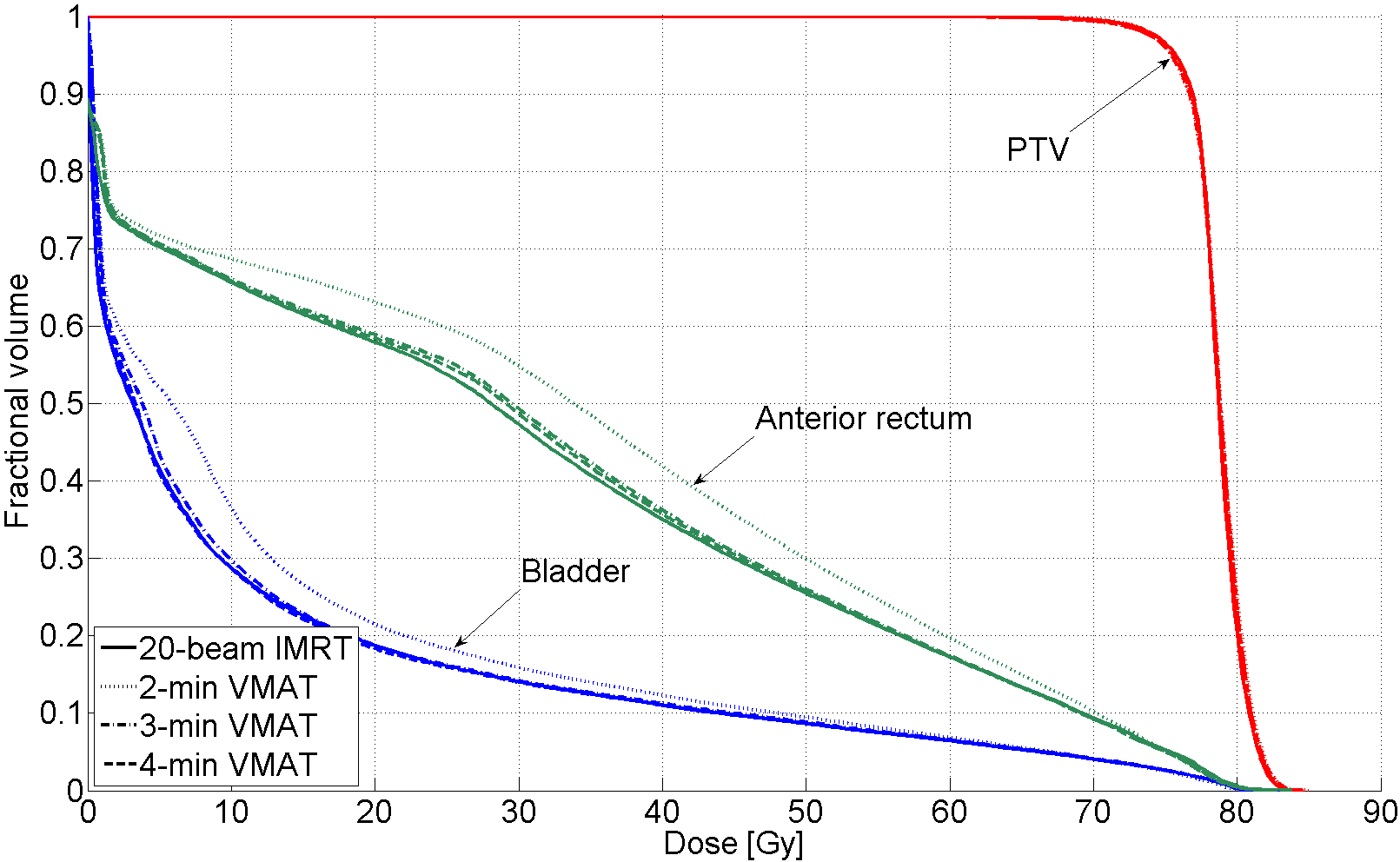}
\caption{Dose-volume histograms for the prostate case comparing an optimal 20-beam IMRT plan and the VMAT plans obtained with treatment time limit $t_{\max}=2$, $3$, and $4$ minutes. VMAT plans optimized for 4 minutes and longer are nearly indistinguishable from the optimal 20-beam IMRT plan. The 20-minute VMAT plan (not shown) is identical to the 20-beam IMRT plan. }\label{fig:prostate-DVH}
\end{figure}

The prostate case has a single target volume that includes both the prostate and the seminal vesicles (prescribed dose $42 \times 1.8 \text{ Gy}$); the dose limiting OARs are the bladder and the anterior rectum. Our experimental setup, including the machine constraints, were identical to the one presented above; we compared an optimal 20-beam IMRT plan to the VMAT plans obtained by our approach for the time limits $t_{\max} = 2, 3, 4$, and $20$ minutes. The DVH curves of the plans are shown on Fig.~\ref{fig:prostate-DVH}.


The results confirm the findings for the head-and-neck case. The 3-minute and 4-minute plans are practically indistinguishable from the optimal IMRT plan in terms of target coverage, conformity, and the sparing of the dose-limiting organs. Also the 2-minute plan is of acceptable quality: the target coverage is identical to that of the IMRT plan, however, it cannot achieve the same OAR sparing. The observation that (in contrast to the head \& neck case) a 3-minute plan yields an almost optimal treatment plan, can be explained by the smaller field size. The smallest radiation field that covers the entire target volume from all angles is $9$ cm wide. This implies that (in total) only one minute is required for the leaves to travel accross the field, whereas two minutes are available for modulation.

\subsection{Paraspinal}

The paraspinal case has a single target volume that surrounds the spinal cord, which is the main dose-limiting organ. The secondary dose-limiting organs in this case are the kidneys, as illustrated in the axial slice shown in Fig.~\ref{fig:hn-plus-paraspinal-colorwashes} (right). The prescribed dose to the PTV is $33 \times 2\text{ Gy}$, the maximum dose in the center of the spinal cord was $50\text{ Gy}$; with a $25\text{ Gy}$ mean dose limit on the kidneys.

\begin{figure}
\centering
\includegraphics[width=400pt]{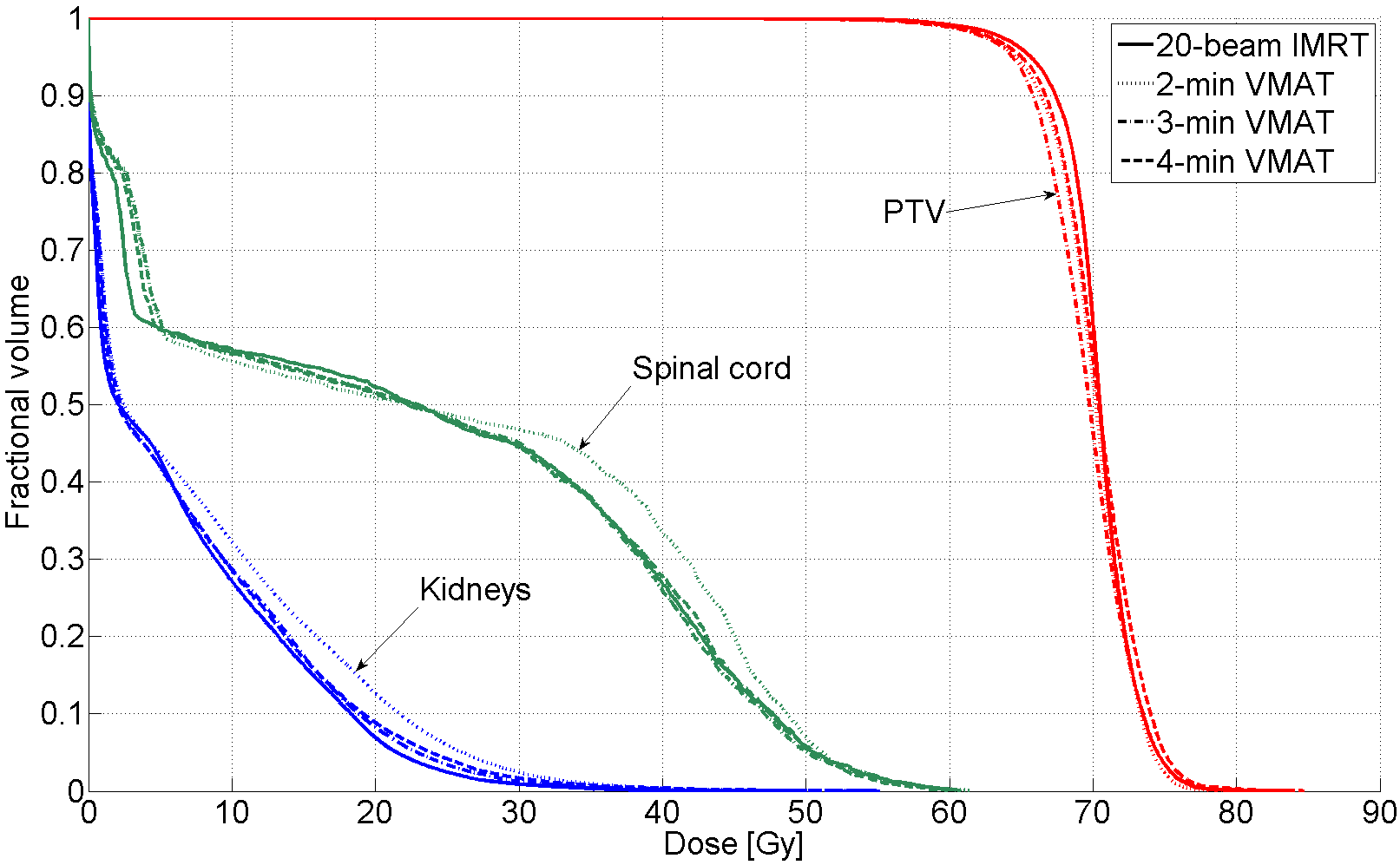}
\caption{Dose-volume histograms for the paraspinal case comparing an optimal 20-beam IMRT plan and the VMAT plans obtained with treatment time limit $t_{\max}=2$, $3$, and $4$ minutes. The 20-minute VMAT plan (not shown) is nearly identical to the 20-beam IMRT plan.}\label{fig:paraspinal-DVH}
\end{figure}

As for the prostate case, we compare an optimal 20-beam IMRT plan to the VMAT plans obtained by our approach for the time limits $t_{\max} =2, 3, 4$ minutes. The DVH curves of the plans are shown on Fig.~\ref{fig:paraspinal-DVH}. The results are consistent with the head \& neck case and the prostate case. Similar to the prostate case, the 2-minute plan is not vastly inferior to the others plans, but the 3-and 4-minute plans achieve better organ sparing. The convergence of the target coverage is slightly slower than for the head \& neck and prostate cases: the 4-minute does not achieve identical target coverage to the 20-beam IMRT plan, but is close. Fig.~\ref{fig:hn-plus-paraspinal-colorwashes} shows the dose distribution of the 4-minute VMAT plan, illustrating the sparing of the spinal cord in the center of the target as well as the avoidance of the kidneys.

\section{Discussion}
\label{sec:discussion}

\paragraph{Sliding window delivery versus arbitrary leaf trajectories:}
A main advantage of the proposed VMAT algorithm consists in formulating VMAT planning as a sequence of convex continuous optimization problem. This recovers one of the central advantages of FMO, i.e. the ability to obtain near optimal solutions. This becomes possible via restricting the leaf trajectories to unidirectional leaf motion. In contrast, attempting to perform direct leaf trajectory optimization while allowing for arbitrary leaf trajectories requires the use of integer variables \cite{craft2009}. Even though arbitrary leaf trajectories have the potential to reduce delivery times further, that comes at the price of solving a much harder optimization problem.



\paragraph{The effect of varying dose deposition matrices over an arc segment:}
As shown in section \ref{sec:model}, the VMAT planning problem is guaranteed to reproduce the ideal FMO solution if 1) the treatment time constraint $t_{max}$ is sufficiently loose and 2) the dose deposition matrices are constant over the arc segment. For the realistic case in which the dose deposition matrices change over the arc segment, the VMAT plan obtained via the algorithm in Section \ref{sec:improved-model} could be slightly better or slightly worse than the FMO plan. However, such differences were not observed in our experiments: For increasing delivery time, the VMAT plan became indistinguishable from the FMO plan regarding the DVHs. This is consistent with earlier findings that using more than 20 beams in IMRT does not significantly improve plan quality (see \cite{bortfeld2010number} and references therein).

\section{Conclusion}
We present a new algorithm for volumetric modulated arc therapy planning that directly optimizes the leaf trajectories of a multi-leaf collimator. In the approach, a 360-degree arc is divided into a given number of arc segments. By enforcing unidirectional leaf motion over an arc segment, the leaf trajectory optimization problem can be solved via a sequence of convex optimization problems. This allows us to reliably determine VMAT plans that are guaranteed to replicate the ideal IMRT dose distribution if the treatment time is sufficiently large. For head \& neck, convergence to the optimal plan quality was observed for 4 minutes delivery time; for prostate and paraspinal tumor geometries 3 minutes delivery time was found to be sufficient.

\section*{Acknowledgements}
The authors are grateful to Drs. Thomas Bortfeld, David Craft, and Ehsan Salari from Massachusetts General Hospital for helpful discussion on the algorithm, and their comments on the manuscript. The research was supported by Philips Radiation Oncology Systems.

\bibliography{references,vmat}

\begin{thebibliography}{10}

\bibitem{yu2011intensity}
Cedric Yu and Grace Tang.
\newblock Intensity-modulated arc therapy: principles, technologies and
  clinical implementation.
\newblock {\em Physics in medicine and biology}, 56(5):R31, 2011.

\bibitem{shepard02}
D.~M. Shepard, M.~A. Earl, X.~A. Li, S.~Naqvi, and C.~Yu.
\newblock Direct aperture optimization: a turnkey solution for step-and-shoot
  {IMRT}.
\newblock {\em Med.\ Phys.}, 29(6):1007--18, 2002.

\bibitem{romeijn05}
H.~E. Romeijn, R.~K. Ahuja, J.~F. Dempsey, and A.~Kumar.
\newblock A column generation approach to radiation therapy treatment planning
  using aperture modulation.
\newblock {\em SIAM Journal on Optimization}, 15(3):838--862, 2005.

\bibitem{carlsson2008combining}
F.~Carlsson.
\newblock Combining segment generation with direct step-and-shoot optimization
  in intensity-modulated radiation therapy.
\newblock {\em Medical physics}, 35:3828, 2008.

\bibitem{cassioli13}
A~Cassioli and J~Unkelbach.
\newblock Aperture shape optimization for {IMRT} treatment planning.
\newblock {\em Physics in Medicine and Biology}, 58(2):301, 2013.

\bibitem{craft09}
D.~Craft and T.~Bortfeld.
\newblock On the tradeoff between treatment time and plan quality in rotational
  arc radiation delivery.
\newblock {\em arXiv preprint arXiv:0910.4934}, 2009.

\bibitem{wang2008arc}
Chao Wang, Shuang Luan, Grace Tang, Danny~Z Chen, Matt~A Earl, and X~Yu Cedric.
\newblock Arc-modulated radiation therapy (amrt): a single-arc form of
  intensity-modulated arc therapy.
\newblock {\em Physics in medicine and biology}, 53(22):6291, 2008.

\bibitem{craft12}
D.~Craft, D.~McQuaid, J.~Wala, W.~Chen, E.~Salari, and T.~Bortfeld.
\newblock Multicriteria {VMAT} optimization.
\newblock {\em Medical physics}, 39(2):686, 2012.

\bibitem{wang2011CCPP}
Danny~Z Chen, Shuang Luan, and Chao Wang.
\newblock Coupled path planning, region optimization, and applications in
  intensity-modulated radiation therapy.
\newblock {\em Algorithmica}, 60(1):152--174, 2011.

\bibitem{Cameron2005sweeping}
C~Cameron.
\newblock Sweeping-window arc therapy: an implementation of rotational {IMRT}
  with automatic beam-weight calculation.
\newblock {\em Phys.\ Med.\ Biol.}, 50(18):4317--4336, 2005.

\bibitem{bzdusek09}
K.~Bzdusek, H.~Friberger, K.~Eriksson, B.~H{\aa}rdemark, D.~Robinson, and
  M.~Kaus.
\newblock Development and evaluation of an efficient approach to volumetric arc
  therapy planning.
\newblock {\em Medical physics}, 36:2328, 2009.

\bibitem{ulrich2007development}
Silke Ulrich, Simeon Nill, and Uwe Oelfke.
\newblock Development of an optimization concept for arc-modulated cone beam
  therapy.
\newblock {\em Physics in medicine and biology}, 52(14):4099, 2007.

\bibitem{otto08}
K.~Otto.
\newblock {Volumetric modulated arc therapy: IMRT in a single gantry arc}.
\newblock {\em Medical physics}, 35:310, 2008.

\bibitem{peng12}
F.~Peng, X.~Jia, X.~Gu, M.A. Epelman, H.E. Romeijn, and S.B. Jiang.
\newblock {A new column-generation-based algorithm for VMAT treatment plan
  optimization}.
\newblock {\em Physics in Medicine and Biology}, 57(14):4569, 2012.

\bibitem{bortfeld2010number}
T.~Bortfeld.
\newblock {The number of beams in IMRT - theoretical investigations and
  implications for single-arc IMRT}.
\newblock {\em Physics in medicine and biology}, 55(1):83, 2010.

\bibitem{deasy2003cerr}
Joseph~O. Deasy, Angel~I. Blanco, and Vanessa~H. Clark.
\newblock {CERR}: a computational environment for radiotherapy research.
\newblock {\em Med.\ Phys.}, 30:979, 2003.

\bibitem{feuvret2006conformity}
Lo{\"\i}c Feuvret, Georges No{\"e}l, Jean-Jacques Mazeron, and Pierre Bey.
\newblock Conformity index: a review.
\newblock {\em Int.\ J.\ Radiat.\ Oncol.\ Biol.\ Phys.}, 64(2):333--342, 2006.

\bibitem{ezzell2003IMRTguidelines}
Gary~A Ezzell, James~M Galvin, Daniel Low, Jatinder~R Palta, Isaac Rosen,
  Michael~B Sharpe, Ping Xia, Ying Xiao, Lei Xing, and Cedric~X Yu.
\newblock Guidance document on delivery, treatment planning, and clinical
  implementation of {IMRT}: Report of the {IMRT} subcommittee of the {AAPM}
  radiation therapy committee.
\newblock {\em Med.\ Phys.}, 30:2089, 2003.

\end{thebibliography}

\end{document}